\documentclass[prb,floatfix,superscriptaddress,citeautoscript,twocolumn]{revtex4-1}
\usepackage{amssymb,amsmath,amsthm,amsfonts}
\usepackage{graphicx}
\usepackage{graphics}
\usepackage{color}
\usepackage{caption}
\usepackage{tabularx}
\usepackage{verbatim}
\usepackage[lofdepth,lotdepth,caption=false]{subfig}
\usepackage[pdfstartview=FitH]{hyperref}
\usepackage[usenames,dvipsnames,svgnames,table]{xcolor}
\usepackage{placeins}
\usepackage{comment}
\usepackage{txfonts} 
\usepackage[cal=cm]{mathalfa} 
\usepackage{cleveref} 
\crefname{equation}{Eq.}{Eqs.}
\crefname{figure}{Fig.}{Figs.}
\usepackage{hyperref}

\newsavebox{\bigleftbox}
\hypersetup{
 colorlinks,
 citecolor=Blue,
 linkcolor=Blue,
 urlcolor=Blue
}

\usepackage{pgffor}
\usepackage{pdfpages}
\makeatletter
\AtBeginDocument{\let\LS@rot\@undefined}
\makeatother

\begin{document}

\title{Structural and Electronic Properties of Defective AlN/GaN Hybrid Nanostrutures}

\author{Ramiro Marcelo dos Santos}
\affiliation{Institute of Physics, University of Bras\'{i}lia , 70.919-970, Bras\'{i}lia, Brazil}
\author{Acrisio Lins de Aguiar} 
\affiliation{Physics Deparment, Federal University of Piau\'{i}, Teresina, Piau\'{i}, 64049-550, Brazil}
\author{Jonathan da Rocha Martins}
\affiliation{Physics Deparment, Federal University of Piau\'{i}, Teresina, Piau\'{i}, 64049-550, Brazil}
\author{Renato Batista dos Santos}
\affiliation{Federal Institute for Education, Science, and Technology Baiano, Senhor do Bonfim, Bahia, 48.970-000, Brazil}
\author{Douglas Soares Galv\~{a}o}
\affiliation{Applied Physics Department and Center for Computing in Engineering and Sciences, State University of Campinas, Campinas, SP, 13083-959, Brazil}
\author{Luiz Antonio Ribeiro Junior}
\affiliation{Institute of Physics, University of Bras\'{i}lia , 70.919-970, Bras\'{i}lia, Brazil}
\date{\today}

\begin{abstract}
Due to the wide range of possible applications, atomically thin two-dimensional heterostructures have attracted much attention. In this work, using first-principles calculations, we investigated the structural and electronic properties of planar AlN/GaN hybrid heterojunctions with the presence of vacancies at their interfaces. Our results reveal that a single vacant site, produced by the removal of Aluminum or Gallium atom, produces similar electronic band structures with localized states within the bandgap. We have also observed a robust magnetic behavior. A nitrogen-vacancy, on the other hand, induces the formation of midgap states with reduced overall magnetization. We have also investigated nanotubes formed by rolling up these heterojunctions. We observed that tube curvature does not substantially affect the electronic and magnetic properties of their parent AlN/GaN heterojunctions. For armchair-like tubes, a transition from direct to indirect bandgap was observed as a consequence of changing the system geometry from 2D towards a quasi-one-dimensional one. The magnetic features presented by the AlN/GaN defective lattices make them good candidates for developing new spintronic technologies.   
\end{abstract}



\maketitle

\section{Introduction}
After the discovery of graphene \cite{novoselov_Science,geim_NM}, several other atomically thin materials with similar structural geometries have been proposed as candidates to active layers of optoelectronic devices \cite{xidong_CSR,manzeli_NRM,miro_CSR}. Graphene is a two-dimensional carbon allotrope with a honeycomb structure with a zero bandgap value. This feature precludes its use in some electronic applications \cite{wang_NP}. To overcome this barrier, other graphene-like monolayers such as h-GaN, h-AlN, h-BN, and MoS$_2$ have recently emerged as promising alternatives for developing new nanotechnologies \cite{xu_CR,shen_NL,morteza_SLM,hussain_RSCA,song_RSCA}. These materials have similar graphene structural properties, but they present different electronic features due to a non-zero bandgap. The electronic structure of homogeneous monolayers composed by graphene-like materials has been widely investigated \cite{palummo_PB,magnusom_PRB,loughin_APL,ruiz_PRB,roland_PRM,tegeler_PSSB,golberg_ACSNANO,kadantsev_SSC,eriksson_PRB} and two-dimensional heterostructures of group-III nitride compounds are still being theoretically and experimentally studied, with important results already obtained \cite{onen_JPCC,onen_PRB,sarigiannidou_IOP,himwas_APL,pawel_AIP,sahin_PRB}. Among these structures, single layers of AlN and GaN have already been synthesized, showing matched lattices and tunable bandgap values \cite{kadioglu_PCCP,kecik_APR,tsipas_APLB,balushi_NM}. In large-scale synthesis process of monolayers, uncontrollable defects such as amorphous solids, vacancy, and contaminants can eventually occur \cite{gonzalez_ASS}. The single-atom vacancy is one of the most known types of lattice defects that can significantly affect the electronic properties of 2D materials  \cite{zhang_INfoMat,lili_JMST,schleberger_MMDPI,wang_CSR,liu_NC}. In structures with partially filled bands, this kind of lattice defects can produce robust magnetic behavior \cite{tucek_CSR,otero_SSR}. 

Besides searching for other two-dimensional graphene-like structures with distinct electronic properties, it is also interesting to look for materials composed of a combination of different kinds of structures similar to graphene, i.e., atomically thin heterostructures. In this sense, it is also possible to obtain heterostructures that present electronic properties different from those exhibited by graphene.  Recently, in-plane composite structures of BN/graphene \cite{sevin_PRB,mengzhu_CM}, GaN/AlN \cite{sahin_PRB,onen_PRB,onen_JPCC}, GaN/SiC \cite{chen_PRB}, and MoS$_2$/WS$_2$ \cite{zhou_RSCA,gong_NM} were theoretically predicted. Particularly, the fabrication of 2D lateral BN/graphene heterostructures has represented a crucial step towards the development of other atomically thin heterostructures that are currently being used in the fabrication of integrated circuits \cite{liu_NATNTECH}. Atomically sharp in-plane heterostructures composed of MoS$_2$/WS$_2$ \cite{gong_NM} and MoS$_2$/WSe$_2$ \cite{liMY_SCIENCE,han_NM}, that present a \textit{p-n} junction signature, have been already fabricated. Moreover, AlN/GaN quantum wells \cite{kladko_JPDAP,sarigiannidou_IOP}, and nanodisks in nanowires \cite{himwas_APL,rigutti_NL,rigutti_PSSA} have also been experimentally investigated as alternative structures to improve photon extraction efficiency in nanodevices. Studies based on first-principles calculations have concluded that in-plane AlN/GaN heterojunctions can be realized to fabricate stable composite heterostructures as thin layers for 2D flexible optoelectronic applications \cite{onen_PRB,onen_PRB2,bacaksiz_PRB,sahin_PRB}. Many efforts have also been devoted to the improvement in obtaining AlGaN/GaN structures due to their promising capabilities for developing such applications \cite{garg_JAP,chung_RP,chang_SCIREP,stanchu_CGD}. However, the understanding of the role played by lattice compositional modes, and interfacial defects, and how this can affect their optoelectronic properties remain unclear even as it is of crucial importance in the development of these composite nanomaterials. 

In this work, we investigated the effects of single-atom vacancies (defects) in hybrid AlN/GaN monolayers and nanotubes. These defects are generated by removing a single aluminum, gallium, or nitrogen atom nearby the interfaces. The calculations are carried out within the framework of Density Functional Theory (DFT) methods, whose computational protocols are detailed in the next section. We have considered free-standing AlN/GaN monolayers since the effects of different substrates on which this system can grow are already known \cite{onen_PRB}. Our results revealed that the magnetic properties of AlN/GaN monolayers and nanotubes are significantly affected by the vacancies. Both 2D and quasi-one-dimensional structures exhibit a magnetic moment as a consequence of these defects. In particular, localized bandgap states are strongly dependent on the type of vacancy. Furthermore, curvature effects do not substantially affect the electronic and magnetic properties of AlN/GaN heterojunctions. Only a transition from direct to the indirect bandgap arises when the system geometry changes from a monolayer towards armchair-like nanotubes. Our results provide further insights into the electronic and structural properties of AlN/GaN heterojunctions that could be exploited in spintronic applications.

\section{Computational Method}
The structural, electronic, and magnetic properties of AlN/GaN monolayers and nanotubes were investigated using a linear combination of atomic orbitals (LCAO)-based DFT approach \cite{hohenberg64,Kohn65} as implemented in the SIESTA code \cite{ordejon96,portal97}. Kohn-Sham orbitals were expanded in a double-$\zeta$ basis set composed of numerical pseudo-atomic spin-polarized orbitals of limited range enhanced with polarization orbitals. The energy shift of 0.02 Ry determines common atomic confinement, which is used to define the cutoff radius for the basis functions. The fineness of the real space grid is determined by a mesh cutoff of 400 Ry \cite{anglada02}. For the exchange-correlation potential, we used the generalized gradient approximation (GGA/PBE) \cite{perdew96}. The pseudopotentials are modeled within the norm-conserving Troullier-Martins \cite{troullier91} scheme, in the Kleinman-Bylander \cite{kleinman82} factorized form. Brillouin-zone integrations were performed using a Monkhorst-Pack \cite{monkhorst76} grid of $15$ $\times$ $15$ $\times$ $1$ ($3$ $\times$ $3$ $\times$ $31$) $k$-points for structural optimization of monolayers (nanotubes). For each structural geometry relaxation, the SCF convergence thresholds for total electronic energy are settled as $10^{-4}$ eV with a density matrix tolerance of 10 $^{-4}$. For AlN/GaN monolayers, periodic boundary conditions are imposed with a perpendicular off-plane lattice vector a$_z$, large enough ($25$ \AA) to prevent spurious interactions between periodic images. In the case of AlN/GaN nanotubes, an off-axis $xy$-vacuum supercell with $30$~\AA{} of box length was used. For all structures, the system converged after the forces on each atom reached the criterion of 0.001 eV/\AA{}.

\section{Armchair AlN/GaN Monolayers}
\begin{figure*}[!htb]
\includegraphics[scale=0.6]{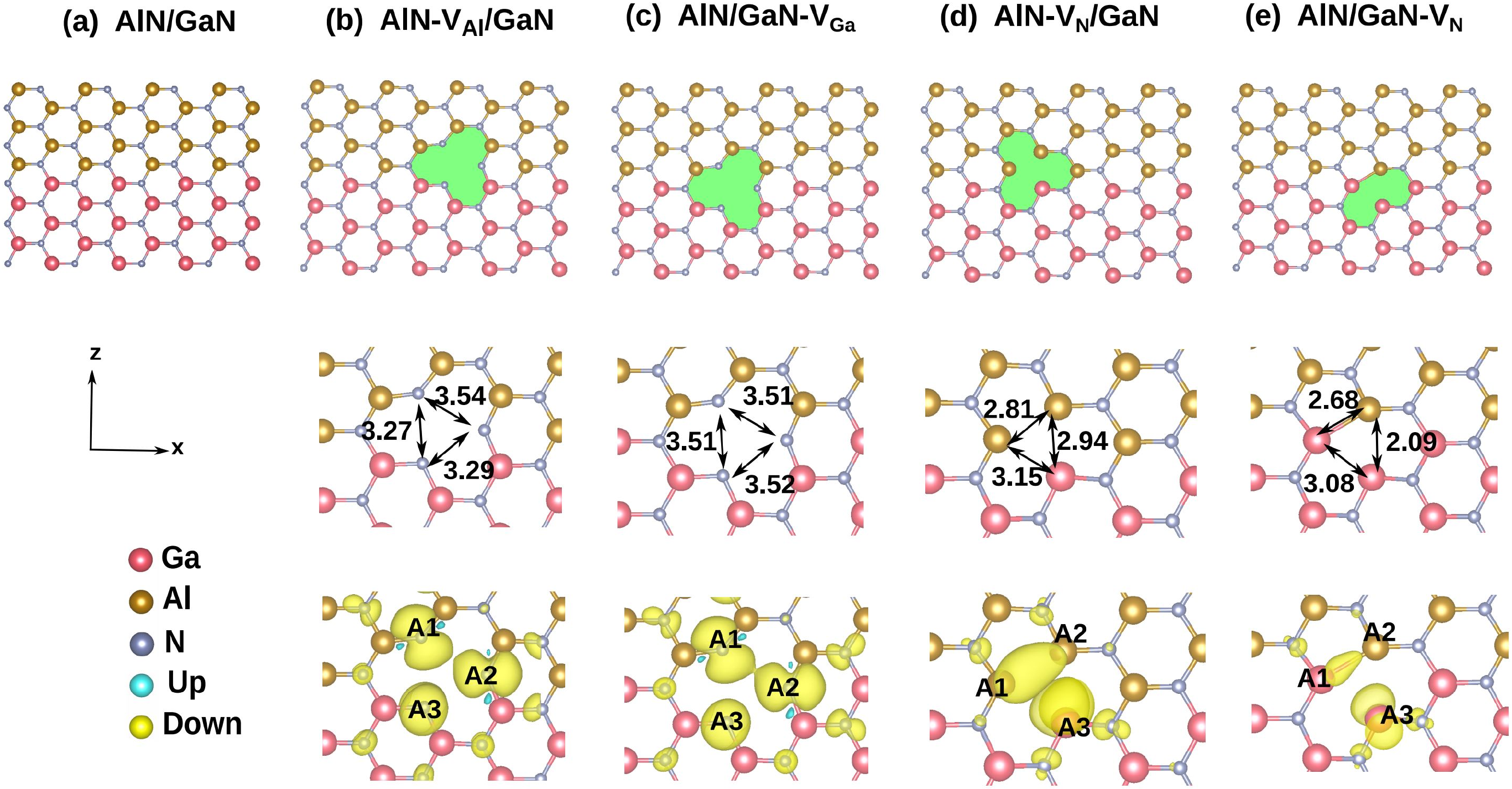}
\caption{\textit{Top panels:} Schematic representation of investigated armchair-like heterojunction monolayers. \textit{Middle panels:} Bond lengths (\AA) values nearby the vacancy region for the corresponding structures shown in the top panels. \textit{Bottom panels:} Charge density population nearby the vacancy region. From left to right, we present the following AlN/GaN heterojunctions: (a) nondefective, (b) with Al-vacancy ($V_{Al}$), (c) with Ga-vacancy ($V_{Ga}$), (d) with N-vacancy in AlN domain (AlN--$V_{N}$), and (e) with N-vacancy in GaN domain (GaN--$V_{N}$). Charge density plots were obtained with isovalues 10$^{-2}$ for V$_{Al}$ and V$_{Ga}$, and 10$^{-3}$ for V$_{N}$. The frontier atoms in the vacancy region are labeled as A1, A2, and A3.}
\label{fig1}
\end{figure*}

We first present a detailed analysis of armchair AlN/GaN monolayers containing vacancies. For comparison purposes, we also investigated zig-zag AlN/GaN monolayers (see Supplementary Material). Armchair structures are known by presenting semiconducting-like transport of quasiparticles. Due to this reason, they are usually of more interest than the zig-zag ones, which normally present only edge-like states \cite{edge1,edge2,edge3}. In Figure \ref{fig1}, we present the supercells of AlN/GaN heterojunctions used in all calculations. Figure \ref{fig1}(a) illustrates the nondefective structure whereas Figures \ref{fig1}(b)-(e) depict the vacancy heterojunctions results. The middle and bottom panels of Figure \ref{fig1} show geometry and charge density configurations for defective layers, respectively. After structural relaxation of the nondefective structure, the obtained average bond length values for Ga--N bond is 1.86 \AA~and for Al--N bond is 1.80 \AA, respectively. These values are in good agreement with bond length distances calculated for pristine h--GaN and h--AlN monolayers \cite{ahangari_SM} and GaN/AlN heterojunctions \cite{sahin_PRB}. According to our calculation, the respective bond lengths for the case of nondefective zig-zag AlN/GaN have the same values shown before for the case of armchair AlN/GaN configuration (See Supplementary Material).

Concerning the structural relaxation of defective layers, we obtained that N--N bond lengths vary from 3.27 up to 3.54 \AA~and from 3.51 up to 3.52 \AA~for AlN-V$_{Al}$/GaN and AlN/GaN-V$_{Ga}$ structures, respectively. These bond lengths are larger than Al--Al (2.81 \AA) and Al--Ga (2.94-3.15 \AA) bond lengths, obtained for AlN-V$_{N}$/GaN defective layers, and also larger than Ga--Ga (3.08 \AA) and Al--Ga (2.68-3.09 \AA) bonds lengths, found for AlN/GaN-V$_{N}$ layers. The structural relaxation procedure employed here yields similar results for defective zigzag AlN/GaN heterojunctions (Supplementary Material). Those small bond lengths found for Al--Al (2.81 \AA) and Ga--Ga (3.08 \AA) in defective sheets, as presented in Figures \ref{fig1}(d) and \ref{fig1}(e), suggests the possibility of a weak covalent bond formation after structural relaxation. For the zigzag AlN/GaN case, those Al--Al and Ga--Ga bond lengths are even smaller, being 2.28 \AA~and 2.56 \AA, respectively (Supplementary Material). On the other hand, for AlN-V$_{Al}$/GaN and AlN/GaN-V$_{Ga}$ cases, since N--N bond lengths increase when compared to nondefective layer, we observe a symmetric and strong repulsion of the electronic cloud nearby the defect. This behavior is similar to one observed by Gonzalez-Ariza and coworkers when studying single-atom Ga and N vacancies in h--GaN layers \cite{gonzalez_ASS}.  

\begin{figure}[!htb]
\includegraphics[width=\linewidth]{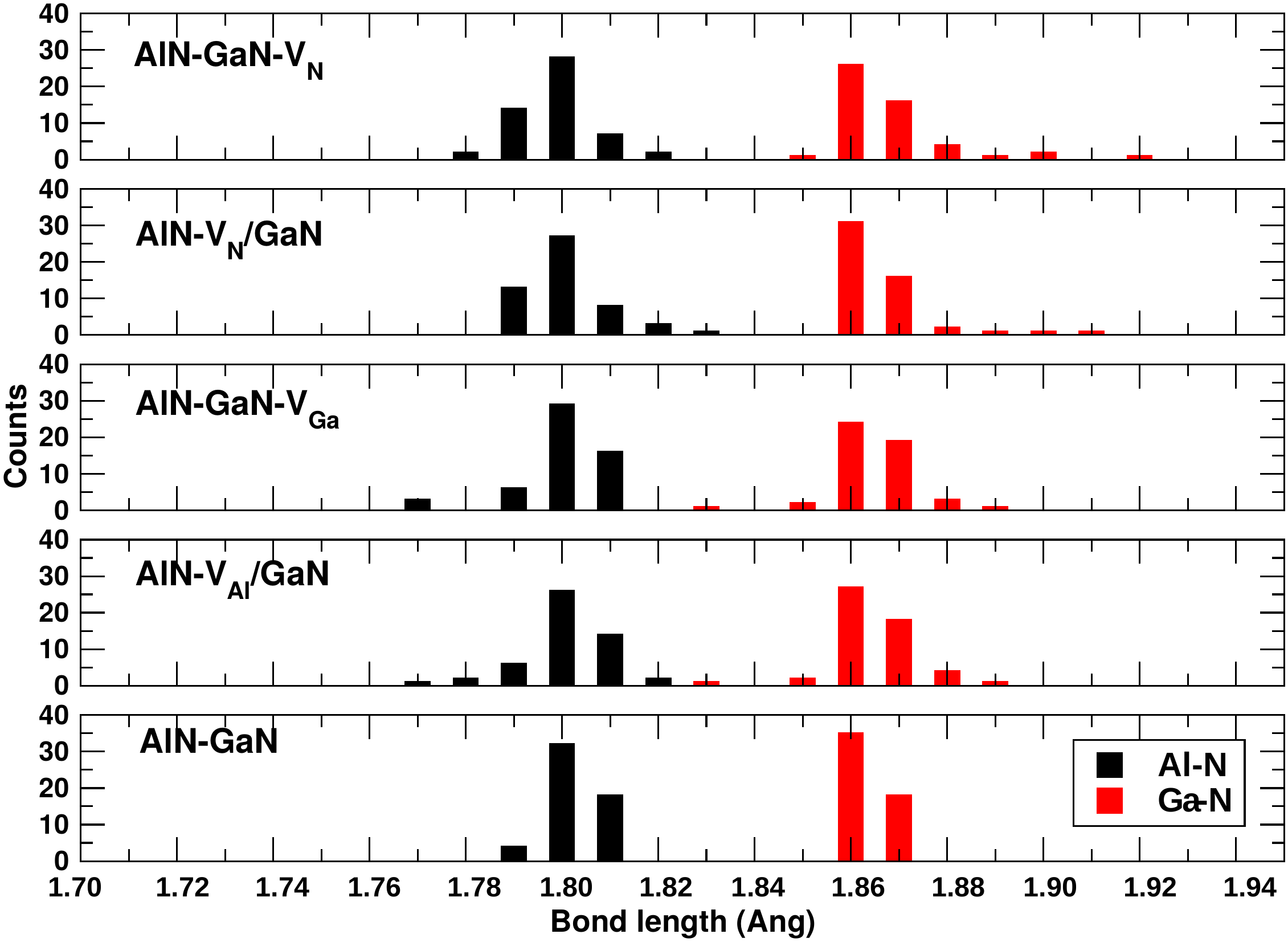}
\caption{Bond length values distribution for the investigated AlN/GaN monolayers. Average values are listed in Table \ref{tab1}.}
\label{fig2}
\end{figure}

Table \ref{tab1} summarizes the structural parameters obtained for AlN/GaN monolayers presented in Figure \ref{fig1}. Besides, we also calculated the distribution of Al--N and Ga--N bond lengths for all investigated systems, as shown in Figure \ref{fig2}. From this figure, one can realize that the minimum and maximum values of Al--N and Ga--N bond lengths in the AlN--V$_{Al}$/GaN case are equivalent to the ones obtained for AlN/GaN--V$_{Ga}$ case. Differences on Ga--N bonds observed in defective layers, when compared to Ga--N bonds of the nondefective case, are more significant for AlN--V$_{N}$/GaN and AlN/GaN--V$_{N}$ cases. These features are mainly due to a substantial bending of GaN domains observed after structural relaxation of the defective nitrogen AlN/GaN layers. Furthermore, we can observe that some Ga and N atoms, close to the vacancy frontier, were slightly shifted along the z-axis direction (out of the plane). It is interesting to observe that AlN domains do not experience such mechanical bendings even for defective nitrogen sheets. Similar results are obtained for zigzag AlN/GaN-V$_{N}$ and AlN-V$_{N}$/GaN layers, which suggest that such mechanical bendings of defective GaN domains are not dependent on the symmetry of AlN/GaN interfaces (see Supplemental Material).

The energetic stability of in-plane AlN/GaN composite structures can be characterized by their average cohesive energies. Table \ref{tab1} also shows the cohesive energy values calculated using the following equation
\begin{equation}
\displaystyle E_{coh} = \lvert E_{tot} - \frac{E_{Al}N_{Al}+E_{Ga}N_{Ga}+E_{N}N_{N}}{N_{Total}} \rvert,  
\end{equation}
\noindent where $E_{tot}$ is the total energy of hybrid (nondefective or defective) monolayer, E$_{X}$ and N$_{X}$ are, respectively, the total energy of isolated X = Ga, Al, or N atoms, and the total number of each X element embedded on each hybrid monolayer. The calculated $E_{coh}$ suggests that AlN/GaN--V$_{N}$ system is the most cohesive heterojunction, followed by AlN--V$_{N}$/GaN structure and both are more cohesive than defective layers with Ga and Al vacancies. Similar conclusions can be made for zigzag AlN/GaN layers (Supplementary Material).

 \begin{table*}[!htb]
 \caption{Structural, electronic, and energetic properties of layered AlN/GaN heterojunctions. Bond distance $d$ (\AA), cohesive energy E$_{coh}$ (eV), energy bandgap E$_{g}$ (eV), magnetic moments $\mu_{B}$, and polarized Mulliken population of the frontier atoms (A1, A2, and A3) are presented for all investigated cases. The indirect bandgap from $\Gamma$ to $X$ point is indicated as IBG, while the direct band gap is referred by DBG in the table.}
 	\begin{tabularx}{\textwidth}{XXXXXX}
     	\hline
                   &   h-AlN/h-GaN        & AlN-V$_{Al}$/GaN & AlN/GaN-V$_{Ga}$  & AlN-V$_{N}$/GaN  & AlN/GaN-V$_{N}$      \\
         \hline
         \hline
          $d_{Al-N}$      & 1.79-1.81           & 1.77-1.82   & 1.77-1.81       & 1.79-1.83   &  1.78-1.82        \\
          $d_{Ga-N}$      & 1.86-1.87           & 1.83-1.88   & 1.83-1.89       &  1.86-1.91    &  1.85-1.92      \\ 
         E$_{coh}$ (eV)   & 4.81          & 4.70          & 4.73           &  4.75          &    4.76            \\
         E$_{g}$ (eV)  &   2.97 (DBG)      &    0.46 (DBG)              &   0.50 (DBG)            &    0.30 (DBG)        &        0.11 (IBG)           \\
      $\mu_{B}$     & 0.00           & 2.71                    &  2.73               &    0.39         &        0.16          \\
 \hline
 A1 (up)     &  -  &       2.063         &      2.067          &      1.346          &        1.384          \\
 A2 (up)     &  -  &       2.083         &      2.078          &      1.403          &        1.428          \\
 A3 (up)     &  -  &       2.216         &      2.215          &      1.335          &        1.375          \\
 \hline     
 A1  (down)  &  -  &       2.924         &       2.922         &      1.483          &         1.409         \\
 A2  (down)  &  -  &       2.938         &       2.938         &      1.462          &         1.448         \\
 A3  (down)  &  -  &       2.935         &       2.940         &      1.475          &         1.442         \\
 	\end{tabularx}
 \label{tab1}
 \end{table*}

Mulliken orbital populations for the vacancy frontier atoms (A1, A2, and A3) can be visualized in the bottom panels of Figure \ref{fig1}. Their values are shown in Table \ref{tab1}. We observe that for AlN--V$_{Al}$/GaN and AlN/GaN--V$_{Ga}$ structures there are unbound $p$--states with approximately the same charge population, where there is a down spin-density majority over vacancy nitrogen atoms. For the AlN--V$_{N}$/GaN and AlN/GaN--V$_{N}$ structures, charge populations are more delocalized over the defect region. It is well-known that nitrogen atoms are mainly found in the gas phase on nature and therefore has no metallic character. This evidence could support the fact that they have a higher charge density localization, characterizing nonbonding states. Conversely, the Al and Ga atoms have a metallic character in their bulk form, which could contribute to a more delocalized charge density. Again, similar orbital populations for the frontier atoms were found for defective zigzag AlN/GaN heterojunctions (Supplementary Material).

\begin{figure}[!htb]
\includegraphics[width=\linewidth]{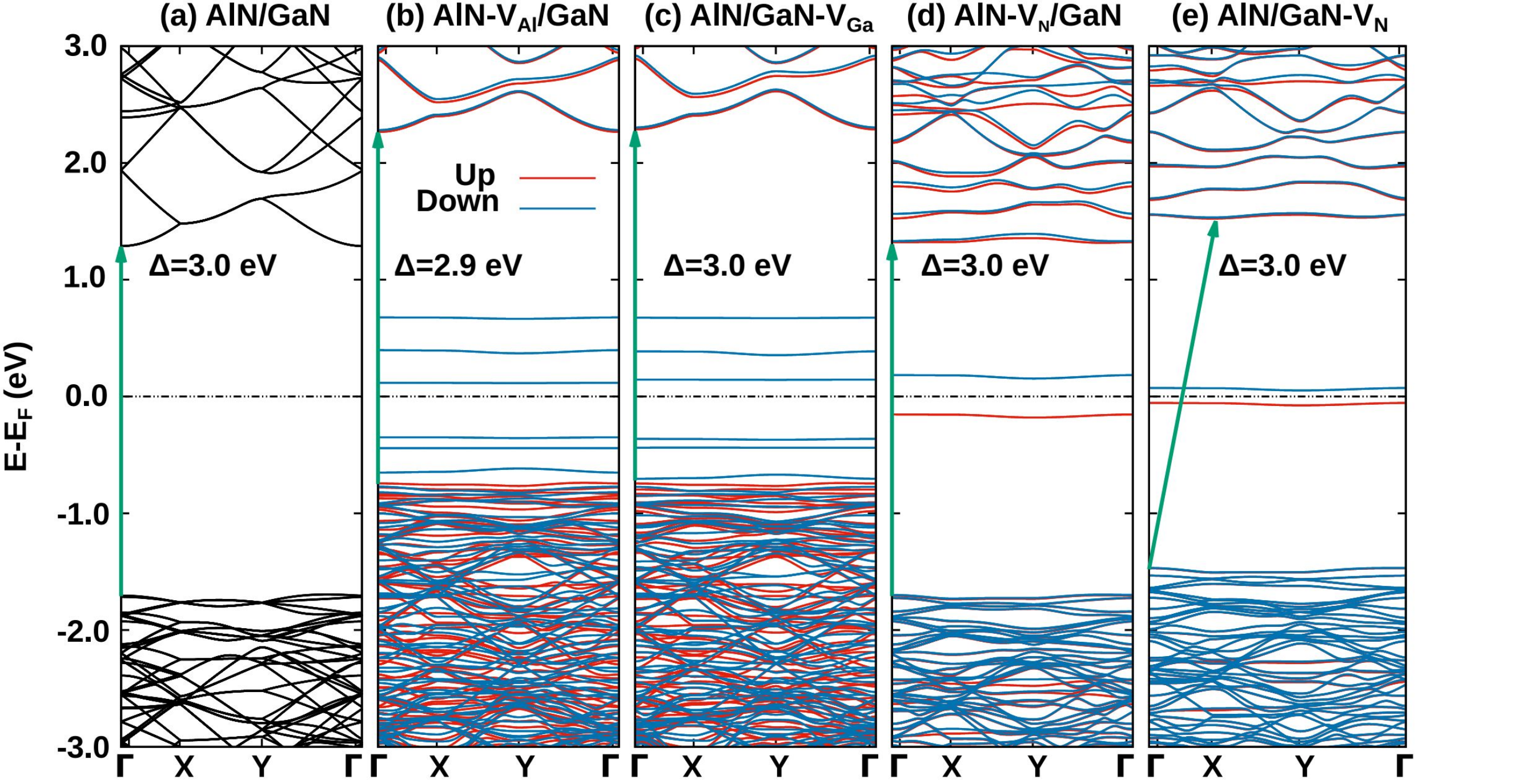}
 \caption{Electronic band structure for (a) nondefective AlN/GaN, (b) AlN--V$_{Al}$/GaN, (c) AlN/GaN--V$_{Ga}$, (d) AlN--V$_{N}$/GaN, and (3) AlN/GaN--V$_{N}$ heterojunctions. Black curves and red arrows indicate, respectively, unpolarized bands and bandgap values.}
 \label{fig3}
\end{figure}

In Figure \ref{fig3}, we present the electronic structure results for the cases discussed above. Figure \ref{fig3} presents electronic band structures for the pristine and defective heterojunctions indicated in Figure \ref{fig1}. The defect-free layer has non-polarized electronic states and exhibits a direct bandgap of 3.0 eV. These results are in agreement with the ones obtained for AlN/GaN heterojunctions by Onen and colleagues \cite{onen_PRB,onen_JPCC}. Zigzag AlN/GaN heterojunctions, in turn, show a slightly smaller direct bandgap (2.8 eV) as reported in the Supplemental Material. In the case of defective heterojunctions, we observed the presence of a strong polarization character --- especially for AlN--V$_{Al}$/GaN and AlN/GaN--V$_{Ga}$ sheets (Figures \ref{fig3}(b) and \ref{fig3}(c), respectively) --- indicated by flat midgap levels along k-path with a spin-down majority. 

\begin{figure}[!htb]
\includegraphics[scale=0.5]{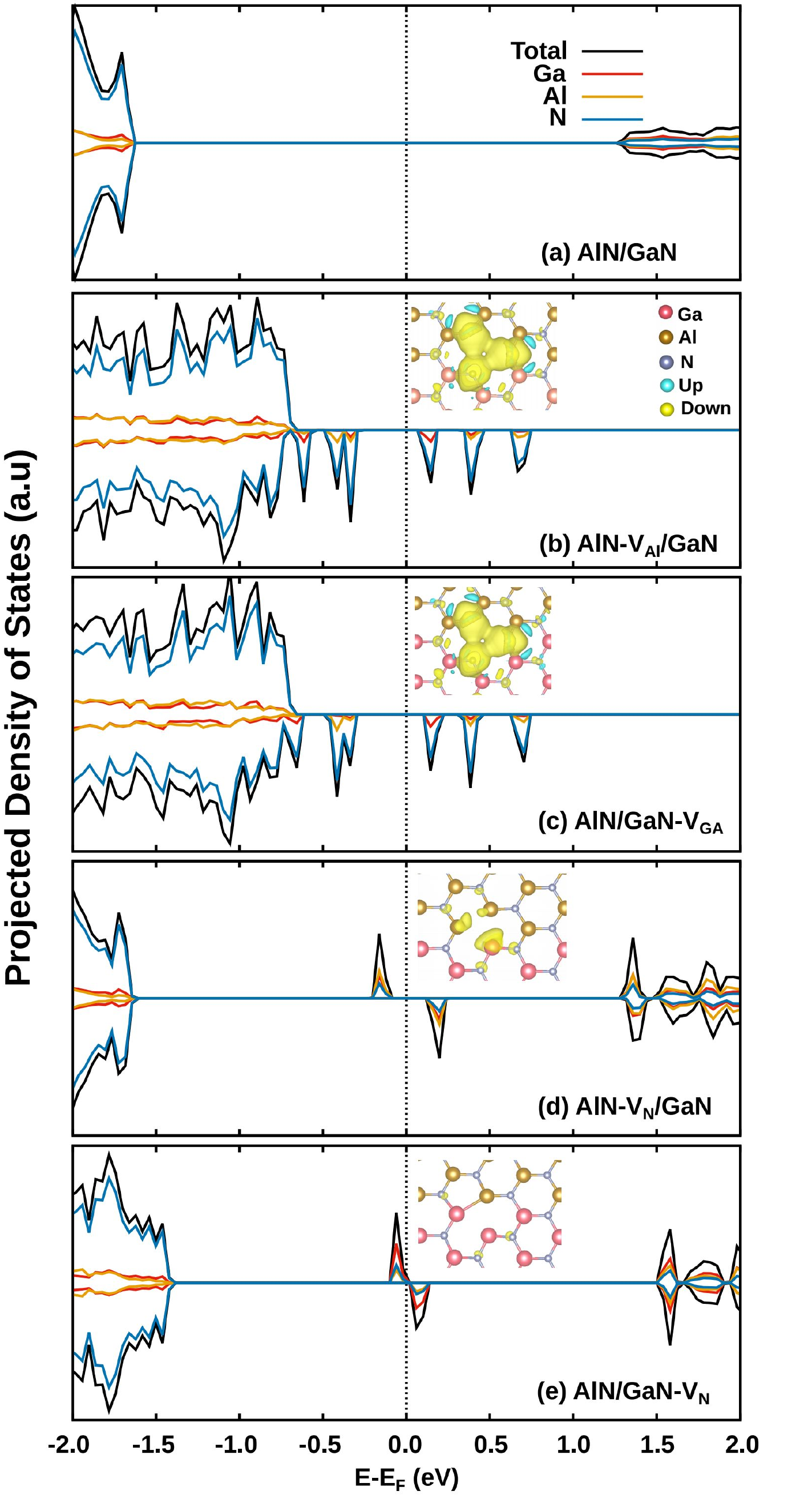}
 \caption{Total (black) and Projected (colored) Density of States (PDOS) calculated for: (a) Aln/GaN, (b) AlN--V$_{Al}$/GaN, (c) AlN/GaN--V$_{Ga}$, (d) AlN--V$_{N}$/GaN, and (e) AlN/GaN--V$_{N}$ monolayers. Inset figures, indicate the corresponding spin density differences between spin up and spin-down ($\rho^{Up} -\rho^{Down}$), zoom-in the vacancy region for each monolayer. LDOS plots were obtained using isovalues of 10$^{-2}$ for V$_{Al}$ and V$_{Ga}$ sheets, and  10$^{-3}$ for V$_{N}$ layers.}
 \label{fig4}
\end{figure}

The calculated total and partial density of states (PDOS) are displayed in Figure \ref{fig4}. As expected and showed in Figure \ref{fig3}(a) and Figure \ref{fig4}(a), the up and down spin densities are symmetric, which leads to a spin density differences between spin-up and spin-down equal to zero. PDOS for Al-vacancy and Ga-vacancy heterojunctions (Figures \ref{fig4}(b) and \ref{fig4}(c), respectively) reveals that those down electronic states are mainly due to the dangling bond of the nitrogen atoms in the center of the defect. In these figures, we can observe that for AlN--V$_{Al}$/GaN and AlN/GaN--V$_{Ga}$ cases, the local charge density is very localized upon nitrogen atoms of the vacancy with a spin-down dominance, as can also be inferred from inset panels that indicate the corresponding spin density differences between spin-up and spin-down ($\rho^{Up} -\rho^{Down}$). These spin excesses are due to the accumulation of dangling states over nitrogen atoms. The absence of Ga or Al atoms in the pristine system breaks the charge distribution symmetry in the heterostructure. From Figures \ref{fig4}(b) and \ref{fig4}(c), we can infer this symmetry breaking by the distribution of spin density. For Ga or Al absence, there is a predominance of spin-down density above the Fermi level. As can be seen in the inserts in Figures \ref{fig4}(b) and \ref{fig4}(c), this predominance came from of the N atoms nearest the vacancy regions. This spin excess contributes for the high magnetic moment values presented in Table\ref{tab1} for AlN--V$_{Al}$/GaN and AlN/GaN--V$_{Ga}$ layers. From the electronic point of view, zigzag AlN--V$_{Al}$/GaN and AlN/GaN--V$_{Ga}$ heterojunctions are similar to armchair ones (see Supplementary Material).

PDOS calculations for systems in the presence of an Al or Ga vacancy (Figures \ref{fig4}(b) and \ref{fig4}(c), respectively) show that these vacancy types contribute to the degeneracy splitting close to Fermi level. The nitrogen-vacancy in the GaN domain generates dangling bonds for two Al atoms. Such bond-breaking promotes the accumulation of spin excess over those elements, as confirmed by PDOS peaks observed for these elements in Figure \ref{fig4}(c). On the other hand, smaller spin excess of N-vacancy AlN/GaN sheets induces small magnetic moment values, as present in Table \ref{tab1}. In this sense, AlN--V$_{Al}$/GaN and AlN/GaN--V$_{Ga}$ lattices present higher magnetic moments than N-vacancy cases. The nitrogen atom has symmetric spherical $S$ and $P$ shells in its two last electronic levels, which significantly reduces the magnetization effects. Ga atoms, in turn, have $d$ and $sp$ orbitals in two last levels, and those orbitals have no spherical symmetry, which contributes to yield higher moment magnetic values obtained for the case shown in Figure \ref{fig4}(c). It is worthwhile to stress that the same behavior is noted to take place when it comes to zigzag AlN/GaN sheets (see Supplementary Material).

Now, analyzing the N-vacancy cases, as shown in Figures \ref{fig4}(d) and \ref{fig4}(e), it is possible to observe the presence of polarized electronic states that are narrowly localized within the bandgap regions for N-vacancy cases. For an N-vacancy in AlN domain (Figure \ref{fig4}(d)), energy levels polarization around Fermi level is characterized for the spin-up channel at the top of the valence band and the spin-down channel in the bottom of the conduction band. Ga atoms have a $4p$-shell with one valence electron, while Al atoms have one valence electron in $3p$-orbitals. This mismatch between orbital momentum values contributes to the polarization of the energy levels around the Fermi level. Similar effects occur for AlN/GaN--V$_{N}$ layer (Figure \ref{fig4}(e)), but, in this case, a vacancy reconstruction takes place due to the formation of an Al--Ga bond. Covalent reconstruction of Al--Al and Ga--Ga bonds, as presented in Figures \ref{fig1}(d) and \ref{fig1}(e), suggests more interaction between those atoms. This behavior could explain the significant symmetric polarization of midgap electronic levels (spin-up and spin-down)  for N-vacancy structures. Similar results were also observed for zigzag AlN/GaN layers. Importantly, such a robust polarization mechanism indicates that N-vacancy AlN/GaN heterojunctions could be desirable systems for future applications into spin-current based devices.

\section{Armchair AlN/GaN Nanotubes}
\begin{center}
\begin{figure*}[htb]
\includegraphics[scale=0.6]{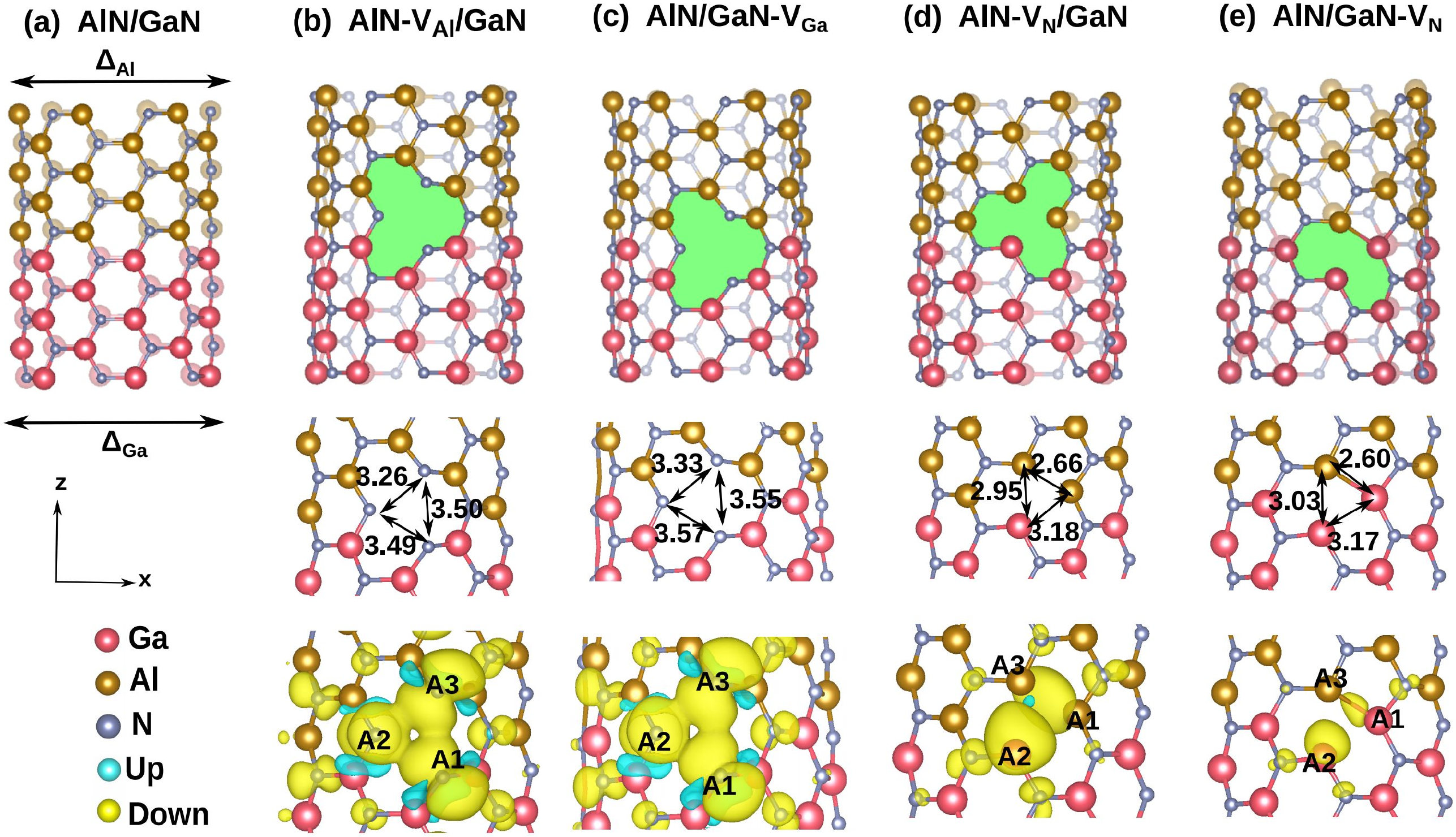}
\caption{\textit{Top panels:} Schematic representation of investigated armchair-like heterojunction nanotubes. \textit{Middle panels:} Bond lengths (\AA) values nearby the vacancy region for the corresponding structures shown in the top panels. \textit{Bottom panels:} Charge density population nearby the vacancy region. From left to right, we present the following AlN/GaN heterojunctions: (a) nondefective, (b) with Al-vacancy ($V_{Al}$), (c) with Ga-vacancy ($V_{Ga}$), (d) with N-vacancy in AlN domain (AlN--$V_{N}$), and (e) with N-vacancy in GaN domain (GaN--$V_{N}$). Charge density plots were obtained with isovalues 10$^{-2}$ for V$_{Al}$ and V$_{Ga}$, and 10$^{-3}$ for V$_{N}$. The frontier atoms in the vacancy region are labeled as A1, A2, and A3.}
\label{fig5}
\end{figure*}
\end{center}

Next, we discuss the previously studied AlN/GaN layers but now rolled up as building blocks to form hypothetical cylindrical tubes. Following the same nomenclature used for monolayers, we model four different armchair AlN/GaN nanotubes containing a monovacancy of Ga, Al, or N nearby their interfaces. These hybrid model nanotubes are presented in Figure \ref{fig5}. For comparison purposes, we also investigated the structural and electronic properties of zigzag AlN/GaN nanotubes (see Supplementary Material). Figure \ref{fig5}(a) shows the schematic representation of a nondefective AlN/GaN nanotube. Since Ga--N and Al--N bonds have slightly different lengths in AlN/GaN layers, different diameters for their nanotube analogs are expected. In Figure \ref{fig5}(a), $\Delta_{Al}$ and $\Delta_{Ga}$ indicate, respectively, the average diameter measured in AlN and GaN domains. Table \ref{tab2} shows the maximum and minimum values obtained for Al--N and Ga--N distances in those heteronanotubes and also the $\Delta_{Al}$ and $\Delta_{Ga}$ values for each system after geometry optimization. In general, Al--N and Ga--N bond lengths for defected heteronanotubes are equivalent to the ones found for their layered analogs, which lead to differences in their average diameters. We obtained that $\Delta_{Al}$ values vary from 10.33 up to 10.37 \AA, while $\Delta_{Ga}$ values range from 10.68 up to 10.71 \AA, which indicates that vacancies do not significantly affect the optimized structure of heteronanotubes.

 \begin{table*}[!htb]
 \caption{Structural, electronic, and energetic properties of AlN/GaN heteronanotubes. Bond distance $d$ (\AA), average nanotube diameters in AlN ($\Delta_{AlN}$) and GaN ($\Delta_{GaN}$) domains, cohesive energy E$_{coh}$ (eV), energy band gap E$_{g}$ (eV), magnetic moments $\mu_{B}$, and polarized Mulliken population of the frontier atoms (A1, A2, and A3) are shown for all investigated AlN/GaN heteronanotubes. All the structures present an indirect band gap from $X$ to $\Gamma$ point.}
 	\begin{tabularx}{\textwidth}{XXXXXX}
     	\hline
                   &   AlN/GaN tube      & AlN-V$_{Al}$/GaN & AlN/GaN-V$_{Ga}$  & AlN-V$_{N}$/GaN  & AlN/GaN-V$_{N}$      \\
         \hline
         \hline
          $d_{Al-N}$      & 1.80-1.80 \AA    &  1.79-1.81 \AA    & 1.79-1.81 \AA   &  1.79-1.83 \AA   & 1.79-1.83 \AA   \\
          $d_{Ga-N}$     & 1.87-1.88 \AA    &  1.86-1.90 \AA    &  1.85-1.90 \AA   &  1.87-1.93 \AA   & 1.86-1.93 \AA   \\ 
   $\Delta_{Al}$ &  10.37 \AA &  10.33 \AA  &  10.37 \AA       &  10.33 \AA & 10.35 \AA \\ 
   $\Delta_{Ga}$ &  10.71 \AA &  10.71 \AA  &  10.68 \AA       &  10.70 \AA & 10.68 \AA \\          
         E$_{coh}$ (eV)   &    4.78 eV       &   4.71 eV         &  4.73 eV         &     4.74 eV      &    4.75 eV  \\
         E$_{g}$ (eV)  &   2.82(a) eV     &    0.69(b) eV             &   0.69(b) eV           &    0.21(c) eV       &     0.04(c) eV             \\
      $\mu_{B}$     &0.0           &2.70                   &   2.69               &    0.30          &      0.06           \\
 \hline
 a (up)     &  -  &   3.017             &   3.018             &    1.334           &        1.373           \\
 b (up)     &  -  &   2.926             &   2.919             &    2.602           &        1.407           \\
 c (up)     &  -   &  2.945              &   2.944             &    1.407           &        1.452           \\
 \hline     
 a  (down)  &  -  &    2.104            &    2.104            &    1.444           &        1.382           \\
 b  (down)  &  -  &    2.186            &    2.185            &    2.607           &        1.407           \\
 c  (down)  &  -  &    2.109            &    2.109            &    1.460           &        1.460           \\
 	\end{tabularx}
 \label{tab2}
 \end{table*}

\begin{figure}[!htb]
 \includegraphics[width=\linewidth]{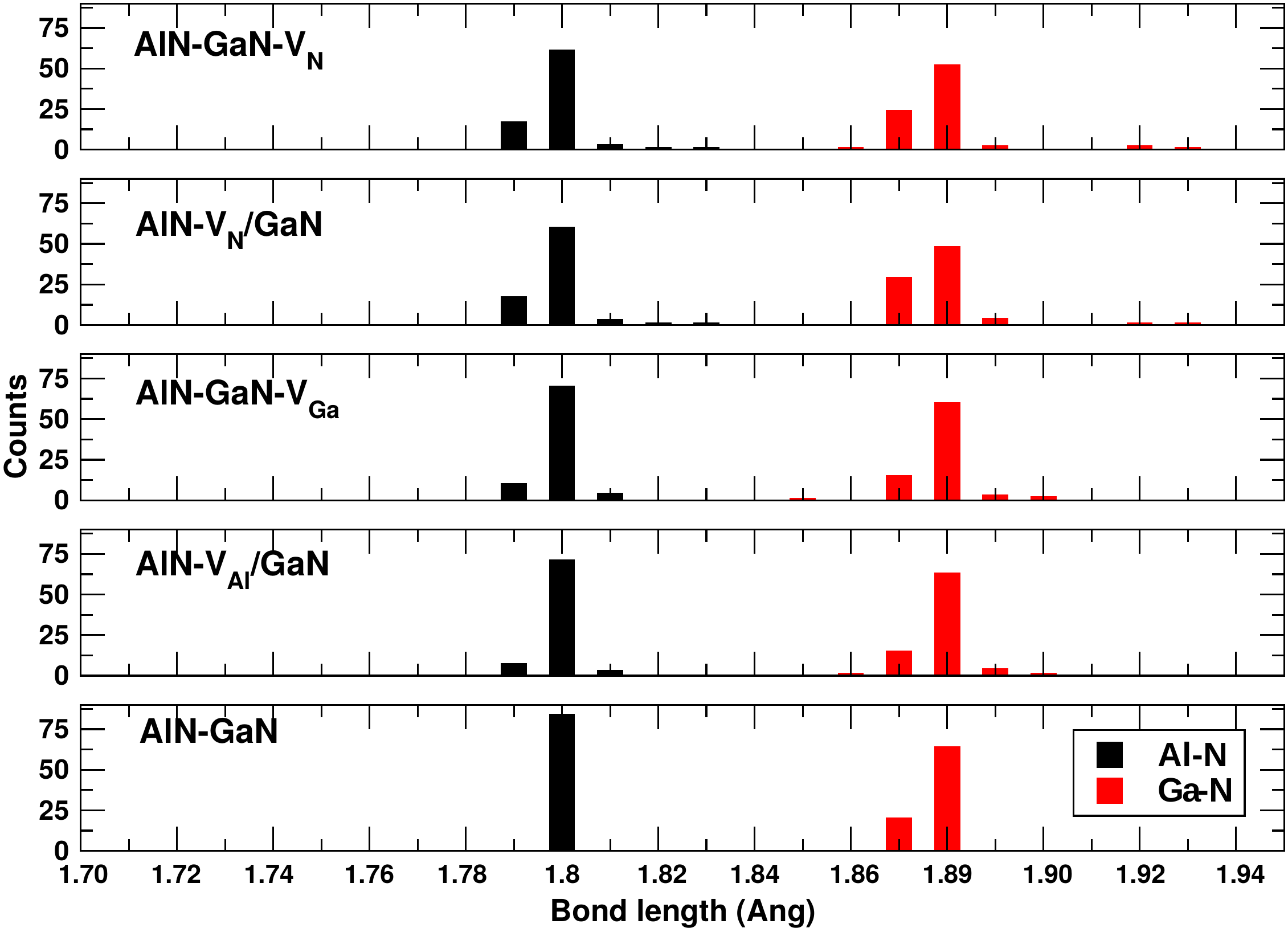}
  \caption{Bond length values distribution for the investigated AlN/GaN nanotubes. Average values are listed in Table \ref{tab2}.}
 \label{fig6}
 \end{figure}

As we can see in Figure \ref{fig6}, the relaxation of defective nanotubes induces N--N bond lengths varying in the intervals 3.26--3.50 \AA~for AlN--V$_{Al}$/GaN case (Figure \ref{fig6}(a)) and  3.33--3.57 \AA~for AlN/GaN--V$_{Ga}$ case (Figure \ref{fig6}(b)), which are similar to their analog layered heterojunctions. These values are larger than Al--Al (2.66 \AA) and Al--Ga bond lengths (2.95-3.18 \AA) found for AlN--V$_{N}$/GaN defective nanotubes (Figure \ref{fig6}(c)), and also larger than Ga--Ga (3.03 \AA) and Al--Ga bond lengths (2.58-3.09 \AA) obtained for AlN/GaN--V$_{N}$ layer (Figure \ref{fig6}(d)). The small bond lengths values found for Al--Al (2.66 \AA) and Al--Ga (2.58 \AA) in defective sheets, as shown in Figure \ref{fig6}(c) and Fig.\ref{fig6}(e), suggest the possibility of a weak covalent bond formation after structural relaxation. Similar to monolayers, for AlN--V$_{Al}$/GaN and AlN/GaN--V$_{Ga}$ nanotube cases, there is a strong and symmetric repulsion of the electronic cloud close to the defect. Figure \ref{fig6} (bottom panels), show the total charge density on vacancy defects. We observe that for AlN--V$_{Al}$/GaN and AlN/GaN--V$_{Ga}$ heteronanotubes, there are unbinding states localized on nitrogen atoms, similar to what was observed for monolayers. In N-vacancy heteronanotube cases, the charge density is delocalized over the nanotube surface. Comparable results for defective zigzag AlN/GaN monolayers were obtained. Moreover, for zigzag AlN/GaN heteronanotubes, those Al--Al and Ga--Ga bond lengths are even smaller, being 2.28 \AA~and 2.56 \AA, respectively.

\begin{figure}[!htb]
\includegraphics[width=\linewidth]{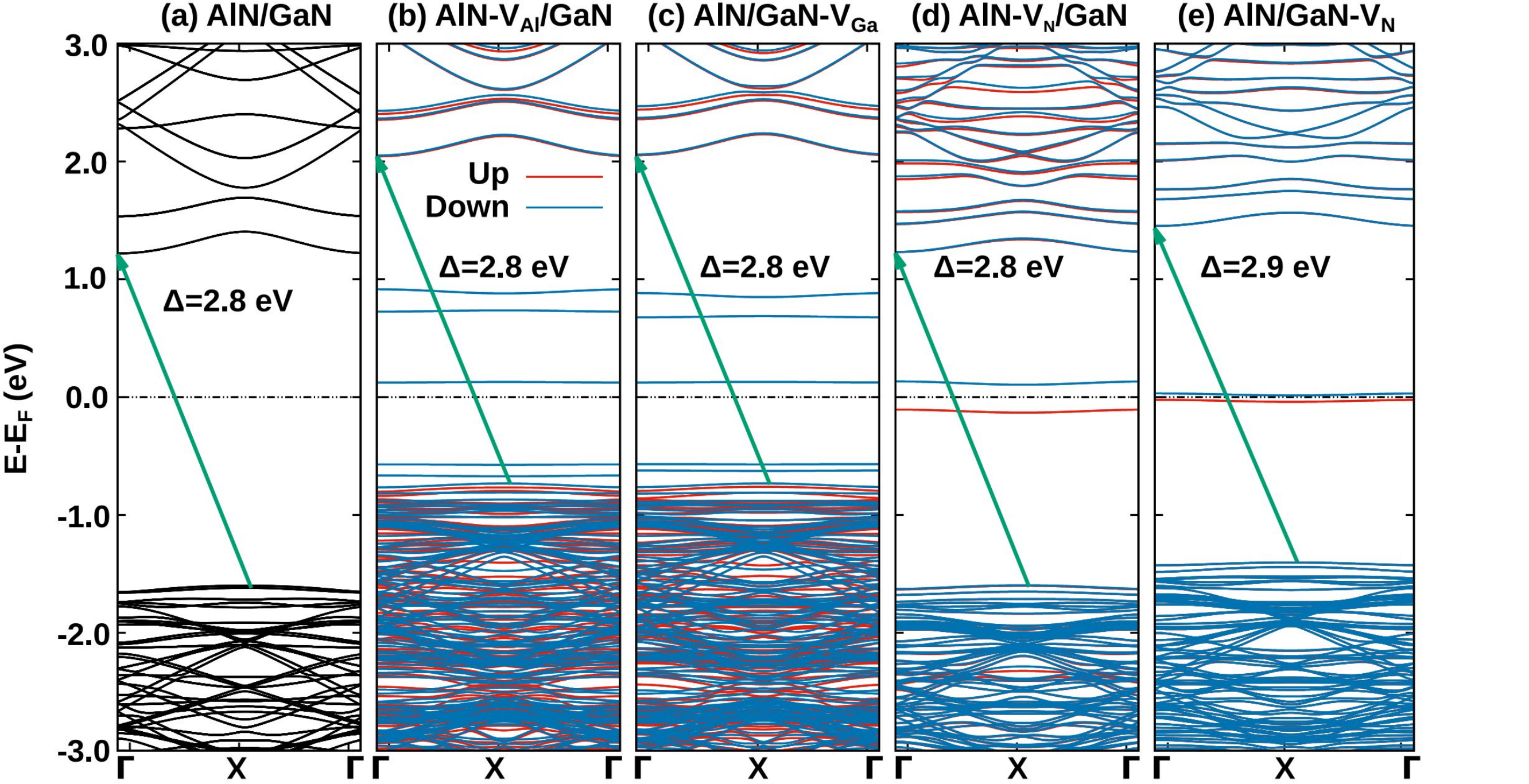}
\caption{Electronic band structure for (a) nondefective AlN/GaN, (b) AlN--V$_{Al}$/GaN, (c) AlN/GaN--V$_{Ga}$, (d) AlN--V$_{N}$/GaN, and (3) and AlN/GaN--V$_{N}$ heteronanotubes. Black curves and red arrows indicate unpolarized bands and bandgaps values, respectively.}
\label{fig7}
\end{figure}

In Figure \ref{fig7}, we present the electronic structure results. Figure \ref{fig7} shows the electronic band structure for all modeled heteronanotubes. Similar to their analog monolayers, heteronanotubes have no polarized electronic states, but with a reduced indirect bandgap, about 2.8 eV from $X$ to $\Gamma$ point. This result suggests that, in general, the curvature effects tend to reduce the bandgap values. Values obtained here for the bandgaps of heteronanotubes are close to the ones obtained by Hui Pan \textit{et. al.} for AlGaN$_2$ nanotubes \cite{huipan_JCTC}. In the case of defective heteronanotubes, we also observe a strong polarization character, especially for AlN--V$_{Al}$/GaN (Figure \ref{fig7}(b)) and AlN/GaN--V$_{Ga}$ (Figure \ref{fig7}(c)) cases, where flat localized levels along the k-path with down majority states are present close to Fermi level. In Figures \ref{fig7}(b) and \ref{fig7}(c), we also observe the presence of a strong polarization character for AlN--V$_{Al}$/GaN (Figure \ref{fig7}(b)) and AlN/GaN--V$_{Ga}$ (Figure \ref{fig7}(c)) cases, similar to what is observed for their monolayer analogs, Figures \ref{fig3}(b) and \ref{fig3}(c), respectively. This lattice polarization is indicated by flat midgap levels along k-path with a spin-down majority when a nitrogen atom is removed from the GaN domain. As discussed above for the monolayer cases, the covalent reconstruction of Al--Al an Al--Ga bonds observed in Figures \ref{fig7}(c) and \ref{fig7}(d) suggests a strong interaction between Al and Ga atoms, which could explain the significant symmetric polarization of midgap electronic levels for N-vacancy structures. 

\begin{figure}[!htb]
\includegraphics[scale=0.5]{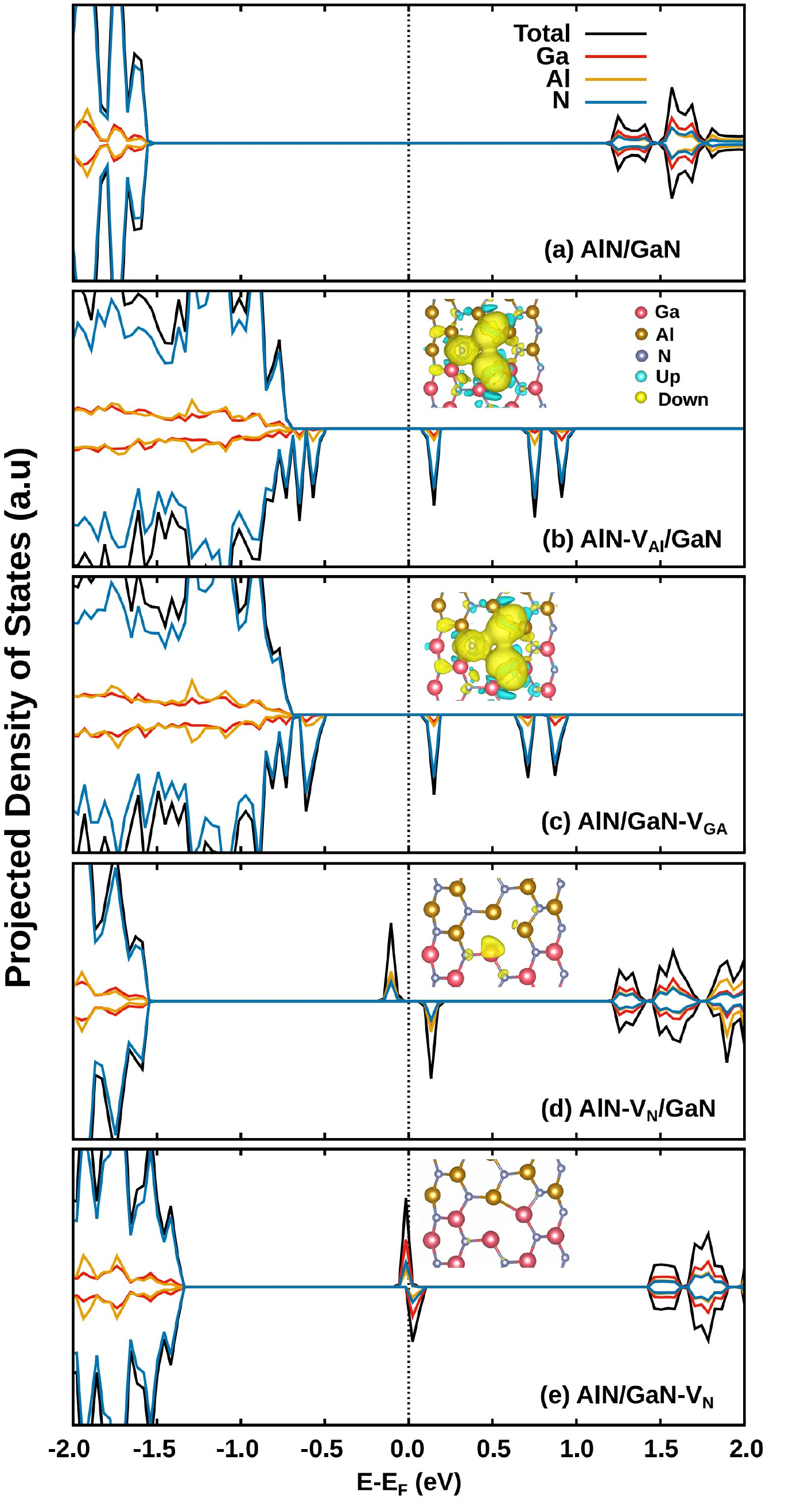}
\caption{Total (black) and Projected (colored) Density of States (PDOS) calculated for: (a) AlN/GaN, (b) AlN--V$_{Al}$/GaN, (c) AlN/GaN--V$_{Ga}$, (d) AlN--V$_{N}$/GaN, and (e) AlN/GaN--V$_{N}$ heteronanotubes. Inset figures indicate the corresponding spin density differences between spin up and spin-down ($\rho^{Up} -\rho^{Down}$), zoom-in the vacancy region for each nanotube. LDOS plots were obtained using isovalues of 10$^{-2}$ for V$_{Al}$ and 10$^{-4}$ V$_{Ga}$, and 10$^{-3}$ for V$_{N}$ in AlN domains and 10$^{-4}$ for V$_{N}$ in GaN domains.}
\label{fig8}
\end{figure}

PDOS calculations for Al-vacancy and Ga-vacancy heteronanotubes are shown in Figures \ref{fig8}(a) and \ref{fig8}(b), respectively. Our results show that those down electronic states are mainly present due to dangling bonds from N atoms nearest to the vacancy region, similar to that was found for their analog monolayer cases. The local charge density, inset panels, confirms the spin excess over Nitrogen atoms. From the electronic point of view, zigzag AlN--V$_{Al}$/GaN and zigzag AlN/GaN--V$_{Ga}$ heterojunctions are similar to armchair ones (see Supplementary Material). Interestingly, all studied zigzag AlN/GaN heteronanotubes present direct band gaps, about 2.88 eV, different of what is presented by armchair ones, where all structures have indirect band gaps. The PDOS calculated for N-vacancy AlN/GaN sheets (Figures \ref{fig8}(c) and \ref{fig8}(d)) reveal that Al, Ga, and also N atoms contribute equally to the up-down spin degeneracy splitting close to Fermi level. The small spin excess in N-vacancy AlN/GaN heteronanotubes (inset panels of Figures \ref{fig8}(c) and \ref{fig8}(d)) yields small magnetic moment values, as presented in Table \ref{tab2}. On the other hand, we can also observe that for AlN--V$_{Al}$/GaN and AlN/GaN--V$_{Ga}$ nanotubes the spin excess (inset panels Figures \ref{fig8}(a) and \ref{fig8}(b)) induces high magnetic moments. Comparable results were also obtained for zigzag AlN/GaN heteronanotubes.

\section{Conclusions}
In summary, we have investigated the structural and electronic properties of defective 2D (monolayer) and quasi-one-dimensional (nanotube) hybrid heterostructures formed by AlN/GaN interfaces. Both kinds of systems show a vacancy reconstruction when it comes to N-vacancy in GaN domains. Moreover, with the presence of vacancies, a significant polarization mechanism was observed. The polarization systems are indicated by the presence of midgap electronic levels, with spin-down for Al and Ga vacancies and spin-up and spin-down for N-vacancy structures. Similar results were also observed for zigzag AlN/GaN in-plane heterojunctions and heteronanotubes. Also, an exciting result found in our calculations is that the pristine armchair AlN/GaN heterostructure has a direct bandgap in its monolayer phase and an indirect bandgap in its nanotube phase. On the contrary, the pristine zigzag AlN/GaN heteronanotube presents a direct bandgap. Even in the presence of vacancy-like defects, the armchair AlN/GaN heteronanotube still has indirect bandgap, but in this last case, with polarization states present in the midgap. Such electronic behavior and robust polarization mechanism indicate that N-vacancy AlN/GaN heterojunctions could be interesting materials for future applications into nanoelectronics technology, mainly in spin-current based devices.

\section*{Acknowledgements}
The authors gratefully acknowledge the financial support from Brazilian research agencies CNPq, CAPES, and FAP-DF. L.A.R.J acknowledges the financial support from a Brazilian Research Council FAP-DF and CNPq grants $00193.0000248/2019-32$ and $302236/2018-0$, respectively. The authors thank CENAPAD-SP and the Laborat\'orio de Simula\c{c}\~ao Computacional Caju\'ina (LSCC) at Universidade Federal do Piau\'i for computational support. A.L.A. acknowledges the brazilian CNPq grant (Process No. $427175/2016-0$) for financial support. DSG thanks the Center for Computing in Engineering and Sciences at Unicamp for financial support through the FAPESP/CEPID Grant \#2013/08293-7.

\bibliography{references}

\foreach \x in {1,...,8}
{
\clearpage
\includepdf[pages={\x}]{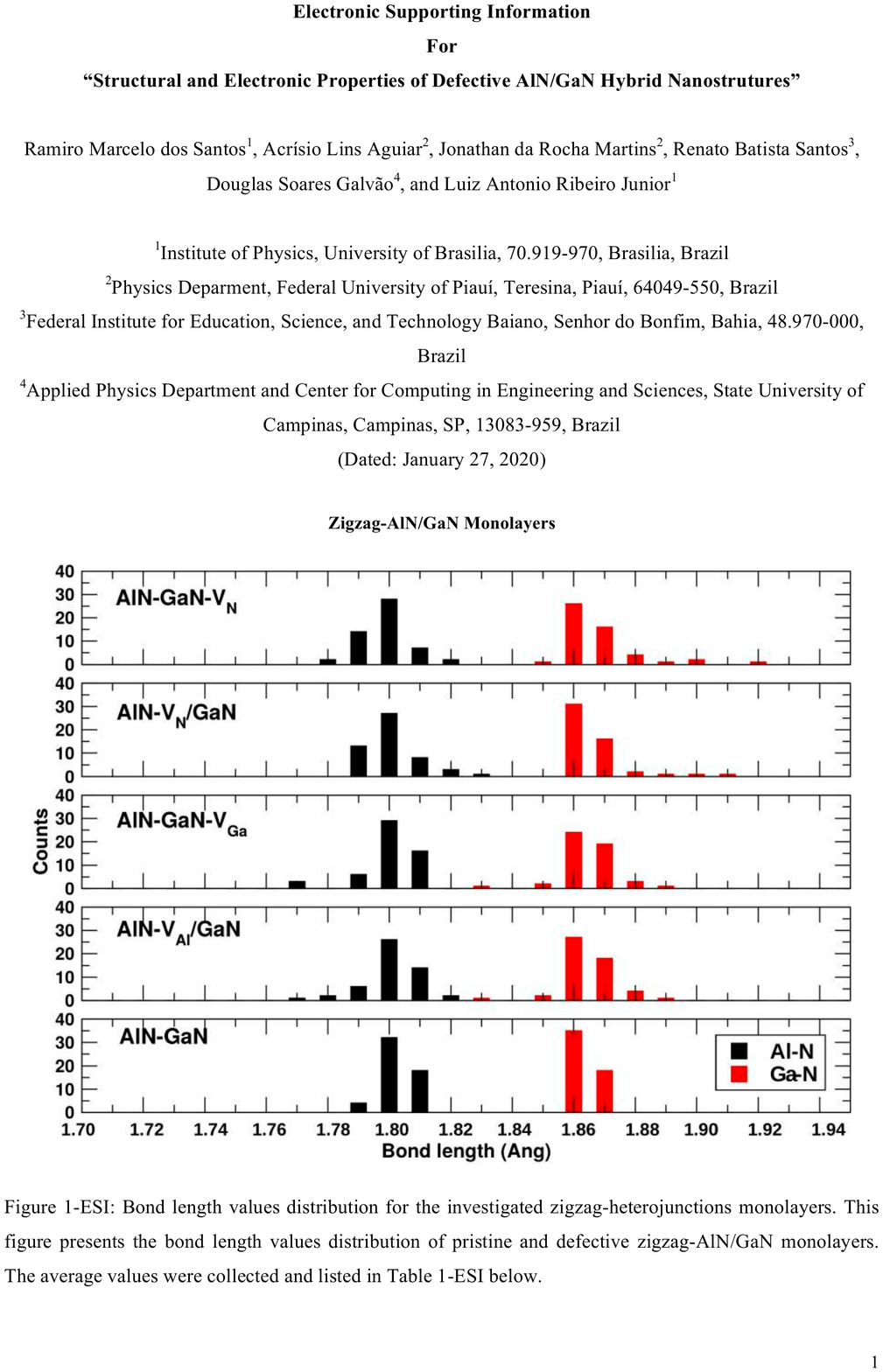} 
}

\end{document}